\documentclass[onecolumn, preprint,amsmath,showpacs,nofootinbib,11pt]{revtex4}
\usepackage{graphicx}
\usepackage{dcolumn}
\usepackage{bm}
\usepackage{subcaption}

\begin{document}

\title{Inflation and de Sitter conjectures in 8-dimensional $R+\gamma R^n$ gravity}

\author{Hai Dang Nguyen}

\affiliation{Faculty of Physics, University of Warsaw, Pasteura 5, 02-093 Warsaw, Poland}

\author{Hoang Nam Cao}

\affiliation{\mbox{Institute of Theoretical and Applied Research, Duy Tan University, Hanoi 100000, Vietnam}}
\affiliation{\mbox{School of Engineering and Technology, Duy Tan University, Da Nang 550000, Vietnam}}

%%%% To generate auto affiliation numbers please use \author{}\affil{} command

\begin{abstract}%
In this work, we study two potentials, the single-field and the two-field, from the modified ($R+\gamma R^n$) gravity in D=8 dimensions. From those potentials, we calculate four observable quantities in inflation, including scalar-to-tensor ratio, spectral index, running index and scalar amplitude. Then, we compare them to the experimental data to verify the righteousness of the models. Last but not least, de Sitter conjectures are brought up with these two potentials to investigate whether it is possible or not the theory lay in the Landscape of quantum gravity. 
\end{abstract}

\maketitle
\section{Introduction}
Theory of General Relativity is a standard explanation for gravity in modern time. In Einstein original statement, gravity depends on one tensor field in linear term, which manifests in the Ricci scalar R of the Einstein-Hilbert action. In fact, gravity can be modified via various ways. The f(R) theory does that by adding one or some non-linear terms of Ricci scalar into the action \cite{sotiriou_fr_2010}. On the other hand, scalar-tensor theory stated that there is a possibility that gravity is not only caused by a tensor field, but a scalar field, which includes kinetics and potential of a Klein-Gordon field into the action \cite{clifton_modified_2012}. Moreover, the two theories can be converted into each other, for example in Starobinsky model, the action of gravity has an extra term $\frac{1}{6M^2}R^2$ beside the Einstein-Hilbert action \cite{starobinsky_new_1980}, then this extra term can be converted into a scalar field with the potential \cite{barrow_premature_1988}:

\begin{equation}
    V= \frac{2}{3}M^2 \left(1-e^{\sqrt{\frac{2}{3}\phi}}\right)^2.
\end{equation}

\subsection{The Model}
Ketov et.al studied the modified gravity, $R+ \gamma R^n - \Lambda$ model, in D-dimensional spacetime \cite{nakada_inflation_2017}. Their idea can be summarized as a process of converting a general D-dimensional f(R) theory into a 4D scalar-tensor theory using conformal transformation and compactification of extra dimensions. For specific, it starts with a general action with an additional $\gamma R^n$ term: \cite{ketov_modified_2019}

\begin{equation} \label{action1}
    S_{grav} = \frac{1}{\kappa^2} \int d^D x \sqrt{-g_D} ( R + \gamma R^n - 2\Lambda ),
\end{equation}
this is an expansion of Starobinsky model with $R^n$ where n is an arbitrary integer, when n=2 it returns Starobinsky model. $\kappa$ is a gravitational coupling constant with dimensions of $\frac{1}{2}(-D+2)$ mass, $\gamma$ is the coupling constant of the higher curvature term and $\Lambda$ is the cosmological constant of D-dimensions.

By Weyl transformation and several steps, an action of a scalar-tensor theory and its two-fields potential in general D dimensions are obtained:

\begin{equation}
    S = \frac{1}{\kappa^2} \int d^D x \sqrt{-\tilde{g}_D} \tilde{R} + \int d^D x \sqrt{-\tilde{g}_D} \left[ -\frac{1}{2}\tilde{g}^{AB} \partial_A \phi \partial_B \phi - V(\phi) \right].
\end{equation}

The first term is the Einstein-Hilbert action in D dimensions, the second term is the kinetic of the scalaron field $\phi$, the last term is the potential:

\begin{multline*}
        \kappa^2 V = \left(\frac{1}{\gamma n} \right)^{\left( {\frac{1}{n-1}} \right)} \frac{n-1}{n} \left( e^{(D-2)\kappa \phi \sqrt{(D-1)(D-2)}}-1 \right)^{\frac{n}{n-1}} e^{-D\kappa \phi /\sqrt{(D-1)(D-2)}}\\
        + 2\Lambda e^{-D\kappa \phi /\sqrt{(D-1)(D-2)}},
\end{multline*}
at this point, the action transformed from an f(R) theory into a scalar-tensor theory with one scalar field $\phi$ that is totally separated from the tensor field.

D cannot be an arbitrary number. Firstly, in the slow-rolls regime, in order for the potential to be plateau when the field goes to infinity, we must have:

\begin{equation}
    n = \frac{D}{2}
\end{equation}

Secondly, n and D-2 must be even, because it is required for the potential to be real when the field is negative. Therefore, D must be multiples of four. Using this conclusion, D=8 was mainly studied, the potential and the action becomes:

\begin{equation}
    \tilde{V}(f) = a^{-2} (1-e^{-6f})^\frac{4}{3} + 2e^{-8f}\tilde{\Lambda}_8.
\end{equation}

\begin{equation} \label{SR8}
     S_8[\tilde{g}_{AB}, f] = \frac{M^6_8}{2} \int d^8 X \sqrt{-\tilde{g_8}} \left( \tilde{R}_8 - 42\tilde{g}^{AB} \partial_A f \partial_B f - M^2_8 \tilde{V}(f) \right),
\end{equation}
with all parameters are in 8-dimensions.

The physical world only has 4 dimensions, thus, compactification is required to make the potential physical. The first two terms of (\ref{SR8}) was compactified with a warp factor $\chi$ \cite{ketov_inflation_2017, otero_r_2017, randall_large_1999}:

\begin{equation}\label{lineelement}
    ds^2_8 = \tilde{g}_{AB}dX^A dX^B =g_{\alpha \beta}dx^\alpha dx^\beta + e^{2\chi} g_{ab}dy^a dy^b,
\end{equation}
the process was performed by compactifying a $(R+R^4)$ modified gravity on a four-sphere $S^4$ with a warp factor $\chi$. After compactification, a new action arrived:

\begin{multline}
    \label{sponcompact}
    S_4[\tilde{g}_{\alpha\beta}, f, \chi] = \frac{M^2_{Pl}}{2}\int d^4x\sqrt{-\hat{g}_4} \left(  \hat{R} - 12  \hat{g}^{\alpha\beta}\partial_\alpha \chi\partial_\beta \chi - 42\hat{g}^{\alpha\beta} \partial_\alpha f \partial_\beta f\right) \\
     -\frac{M^2_{Pl}}{2}\int d^4x\sqrt{-\hat{g}_4} \left( e^{-4\chi} M^2_{Pl}\left[\tilde{V}(f) - 2e^{-2\chi} \right] \right),
\end{multline}
this effective action has two fields, which means a new moduli $\chi$ emerged from the process. This moduli has its own kinetic but it also couples to the inflaton f potential therefore it would interupt the dynamics of the inflation if its mass was lower than the inflationary scale. There, it needs to be stabilized later.

Next, the potential was compactified with Freund-Rubin method, which is usually called flux compactification, with a p-form gauge where p = n \cite{ketov_modified_2019, douglas_flux_2007}. As a result, Ketov derived a potential in D = 4 spactimes with two fields: a radion field $\chi$ and an inflaton field f:
\begin{equation} \label{2fieldspotential}
    M^{-2}_{Pl} V(\chi,f) = \left[ a^{-2}(1-e^{-6f})^\frac{4}{3} + 2\tilde{\Lambda}_8 e^{-8f} \right] e^{-4\chi} - 2e^{-6\chi} + F^2 e^{-12\chi},
\end{equation}
 with three parameters $a^{-2}$, $F^2$ and $\tilde{\Lambda}_8$

\begin{figure}[h]
    \centering
    \includegraphics[width=10cm]{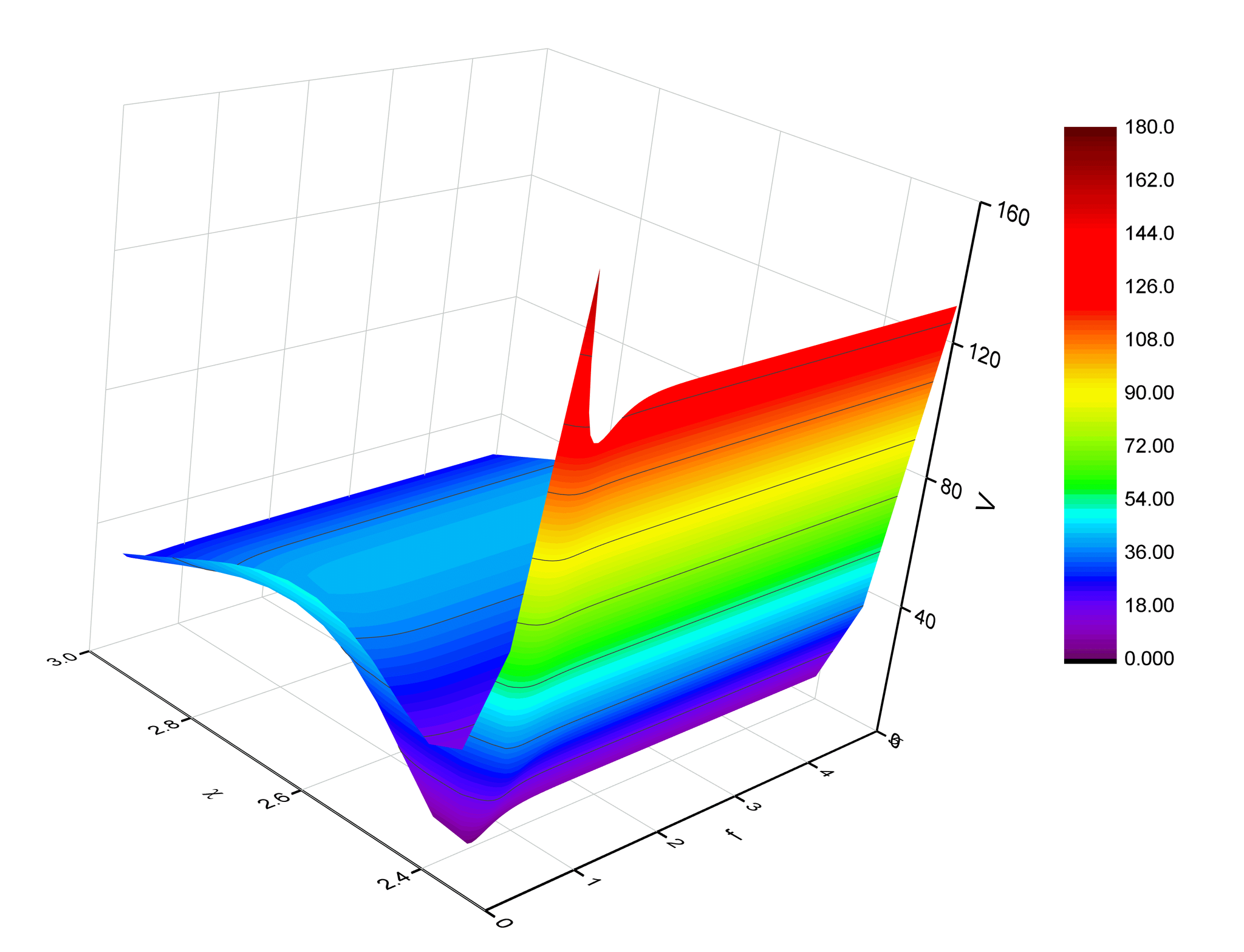}
    \caption{Two-fields potential (\ref{2fieldspotential}) with a = 9.1027, $F^2 = 10^6$ and $\tilde{\Lambda}_8 = 0.01$.}
    \label{fig:my_label}
\end{figure}

Lastly, stabilization of the moduli $\chi$ is required. As a result, a single-field potential with just one inflaton field f is obtained:
\begin{equation}\label{singlefieldV}
    e^{4\chi_0}a^2 V(f) = \left(1-e^{-6f} \right)^\frac{4}{3} + \lambda e^{-8f}-\lambda\left(1+\lambda^3\right)^{-\frac{1}{3}},
\end{equation}
with only one parameter $\lambda = 2a^2 \tilde{\Lambda}_8$. \cite{ketov_modified_2019}

\subsection{Inflation}
The Big Bang model gave rise to multiple of problems including the horizontal problem and flatness problem which can be solved by a rapidly expansion phase of the early universe called inflation. Inflation involves a real scalar field called inflaton $\phi$, the dynamics of this field obey the Klein-Gordon equations with a general potential V($\phi$): \cite{linde_new_1982, linde_axions_1991, linde_chaotic_1983, linde_current_2005}

\begin{equation}\label{eom}
    \ddot{\phi} +\frac{\partial V}{\partial \phi} + 3H\phi = 0
\end{equation}
where $H = \frac{\dot{a}}{a}$ is the Hubble constant, a is scale factor. The scale factor relates to an important quantity called the e-folding number N(t) by $a=a_{end}e^{N(t)}$ where $a_{end}$ is the scale factor at the end of inflation era. The equation of motion \ref{eom} is usually studied with slow-roll approximation with the assumption the potential of the field dominates over its kinetic. Depends on the potential $V(\phi)$, there are several models for inflation. Three most usually studied models are the large field, small field and hybrid. \cite{linde_current_2005}

\subsection{The Swampland Conjecture}

The Swampland program provided a strong connection between quantum gravity in ultraviolet and effective field theories. In the string Swampland, the set of all consistent effective field theories is divided into two regions: the Landscape, where these theories are potential quantum gravity theory in UV; the Swampland, where contains theories that cannot do the same. There are various conjectures, including distance conjecture, weak gravity conjecture, etc. In this work, de Sitter conjectures are considered \cite{vafa_string_2005, palti_swampland_2019}.

An inflaton field f action in D-dimensional spacetime with f(R) theory is
\begin{equation}
    S = \int dx^D \sqrt{-g_D} \left( \frac{1}{2}M_{Pl} f(R_D) -\frac{1}{2}g^{AB}\partial_A f \partial_B f - V(f) \right),
\end{equation}

where $M_p$ is the Planck mass, AB are indices in D-dimensions, V is the potential of the inflaton field. Based on string theory, the original de Sitter conjecture states that \cite{palti_swampland_2019, danielsson_what_2018}

\begin{equation}
    \left| \nabla V \right| \geq \frac{c}{M_{Pl}}V.
\end{equation}

Moreover, if one consider entropy of de Sitter space, the refined de Sitter conjecture adds \cite{palti_swampland_2019, ooguri_distance_2019}

\begin{equation}
    min\left(\nabla_i \nabla_j\right) \leq -\frac{c'}{M_{Pl}^2} V,
\end{equation}

with both c and c' must be positive constants of order 1. Furthermore, Andriot et. al in 2019 proposed a further refined conjecture: \cite{andriot_further_2019, liu_higgs_2021}

\begin{align}
    \label{furrefdeSitter}
    \left( M_p \frac{|\nabla V|}{V} \right)^q -aM^2_p \frac{min(\nabla_i \nabla_j)}{V} \geq b,
\end{align}

with positive a+b=1 and $q>2$.

\section{Inflation in $R+\gamma R^n$ gravity}

\subsection{Inflation with a single-field potential}

The inflationary with one scalar field minimally coupled to gravity in D-dimension has following action:
\begin{equation}
    \mathcal{S} = \int d^Dx \sqrt{-g_D} \left[ \frac{M_{Pl}^2}{2}R_D - \frac{1}{2}\partial_A\partial^A f - V(f) \right],
\end{equation}
with $R_D$ is the Ricci scalar in D-dimension and $M_{Pl}=(8\pi G)^{-1/2}$ is the reduced Planck mass. The potential V(f) is defined in \ref{singlefieldV}.

The calculation started with three slow-rolls parameters $\epsilon$, $\eta$ and $\zeta^2$, relates to the single-field potential by these relations: \cite{nam_implications_2021}
\begin{equation}
    \begin{split}
        \epsilon = \frac{M_P^2}{2} \left( \frac{V'}{V} \right)^2, \\
        \eta = M_P^2 \frac{V''}{V},\\
        \zeta^2 = M_P^4 \frac{V^{(3)}V'}{V^2}.
    \end{split}
\end{equation}

The value of the field is calculated from the e-foldings number:

\begin{equation}
    N = \int^{f_{in}}_{f_{end}} \frac{1}{M^2_{Pl}}\frac{V}{V'(f)}df,
\end{equation}
where $f_{in}$ will be the value of the field, which is the horizon crossing value, $f_{end}$ is the value of inflation when it ends $\epsilon(f_{end})\approx 1$. For the results from N=58 to N=62, they are given in \ref{tab:onefield} with $\lambda = 0.01$.

\begin{table}[h]
    \centering
    \caption{Four observable quantities in each e-folding from 58-62.}
\begin{tabular}{|c|c|c|c|c|c|}
    \hline
    $N_e$ & $f$ & $n_s$ & r & $\frac{dn_s}{d ln k}$ & $A_s$\\
    \hline \hline
   58	&4.62035Mp	&0.966845	&0.00245153	&-5.56983 $\times 10^{-4}$	&4.49033$\times10^{-8}$\\
\hline
59	&4.63785Mp	&0.967398	&0.00237233	&-5.38474$\times 10^{-4}$	&4.64167$\times10^{-8}$\\
\hline
60	&4.6547Mp	&0.96792	&0.00229852	&-5.21248$\times 10^{-4}$	&4.7921 $\times10^{-8}$\\
\hline
61	&4.67156Mp	&0.968434	&0.00222699	&-5.04577 $\times 10^{-4}$	&4.94742$\times10^{-8}$\\
\hline
62	&4.68841Mp	&0.968939	&0.00215773	&-4.88456$\times 10^{-4}$	&5.10765$\times10^{-8}$\\
\hline
    \end{tabular}   
    \label{tab:onefield}
\end{table}

Generally, the predicted values of these quantities depends on the parameter $\lambda$. Therefore, the dependence of the observable quantities on the cosmological constant - $\lambda$ from -1 to 3 is:
\begin{figure}[h] 
    \begin{subfigure}{.47\textwidth}
    \centering
    \includegraphics[width=.8\linewidth]{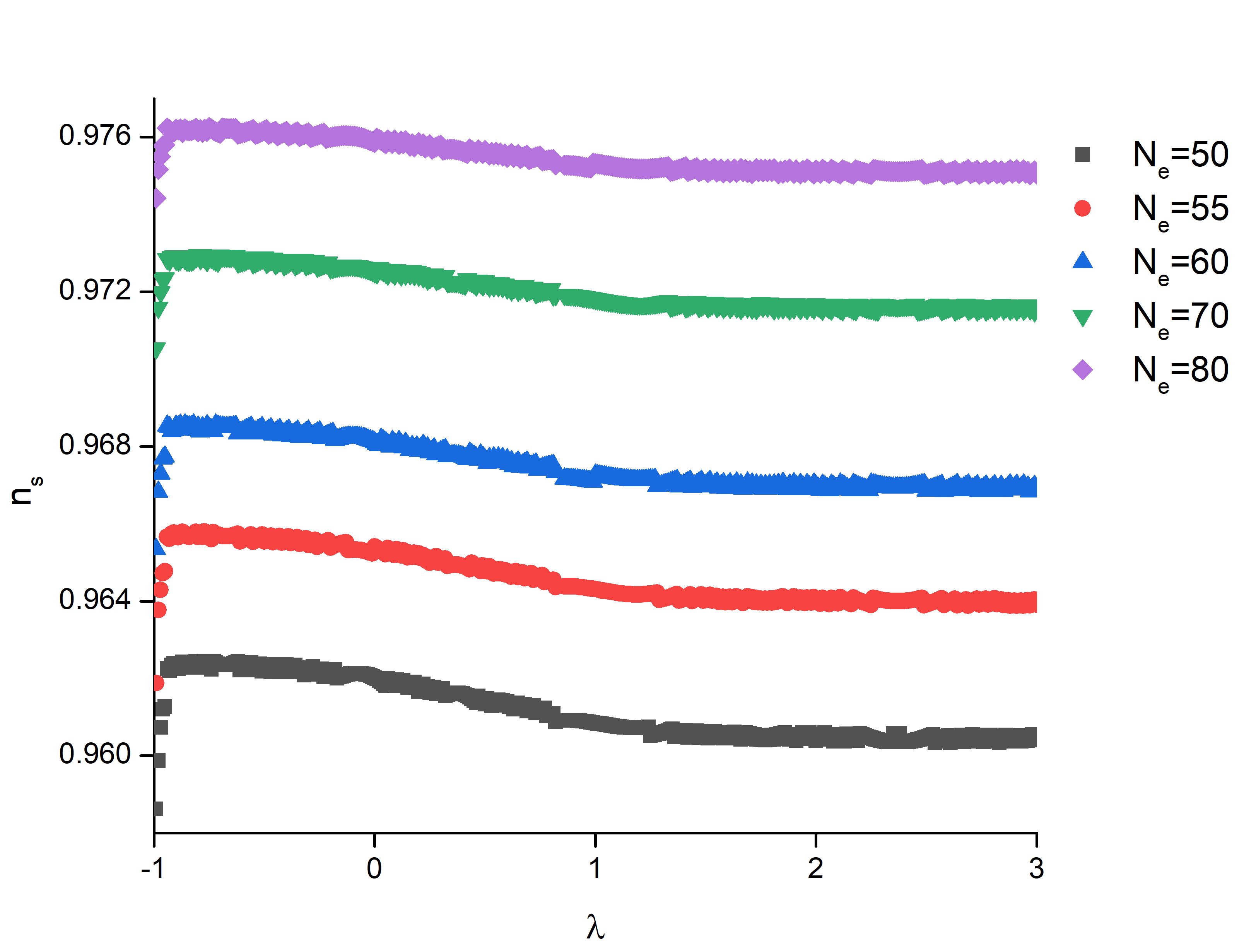}
    \caption{}
    \label{ns}
\end{subfigure}
\begin{subfigure}{.47\textwidth}
    \centering
    \includegraphics[width=.8\linewidth]{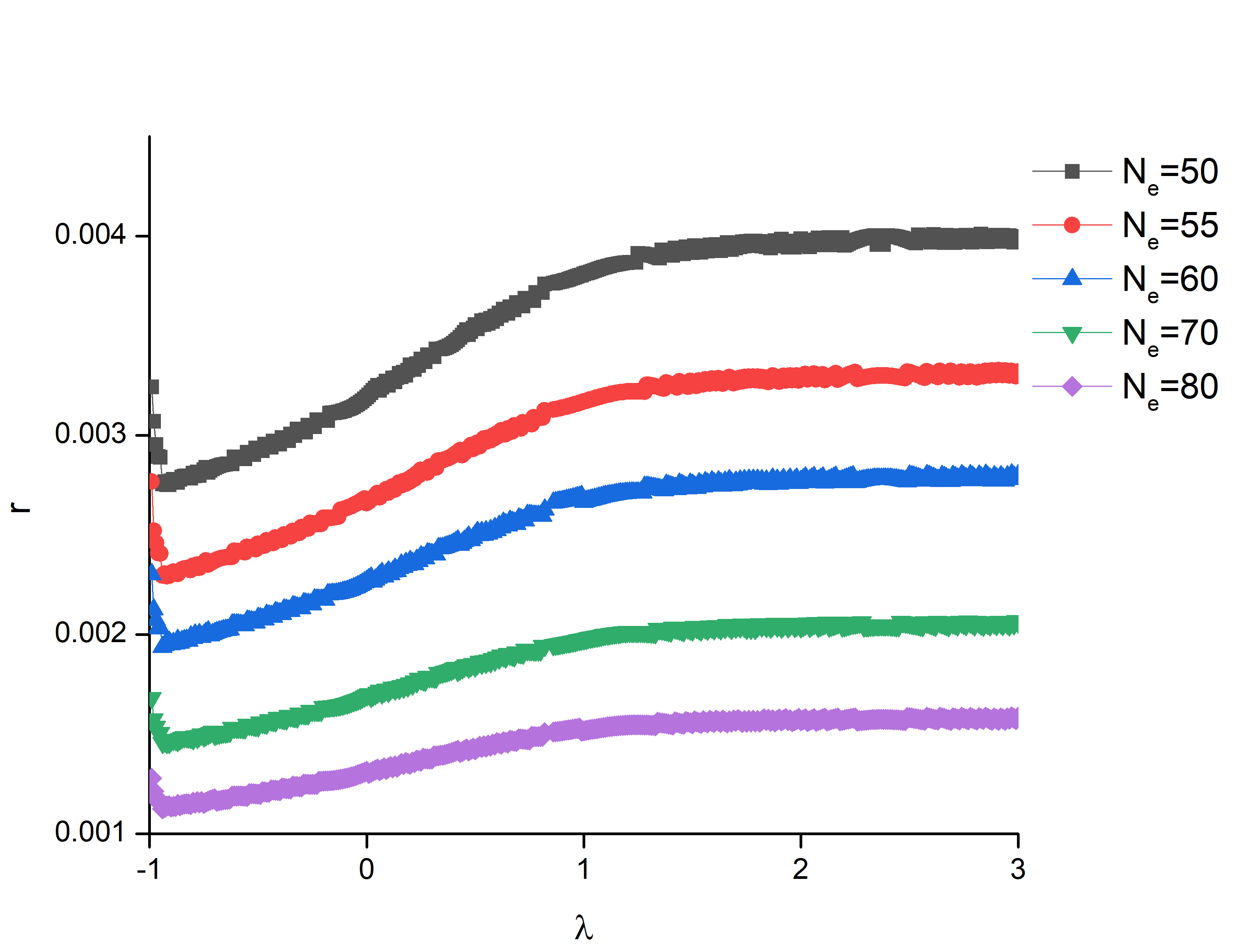}
    \caption{}
    \label{r}
\end{subfigure}
\begin{subfigure}{.47\textwidth}
    \centering
    \includegraphics[width=.8\linewidth]{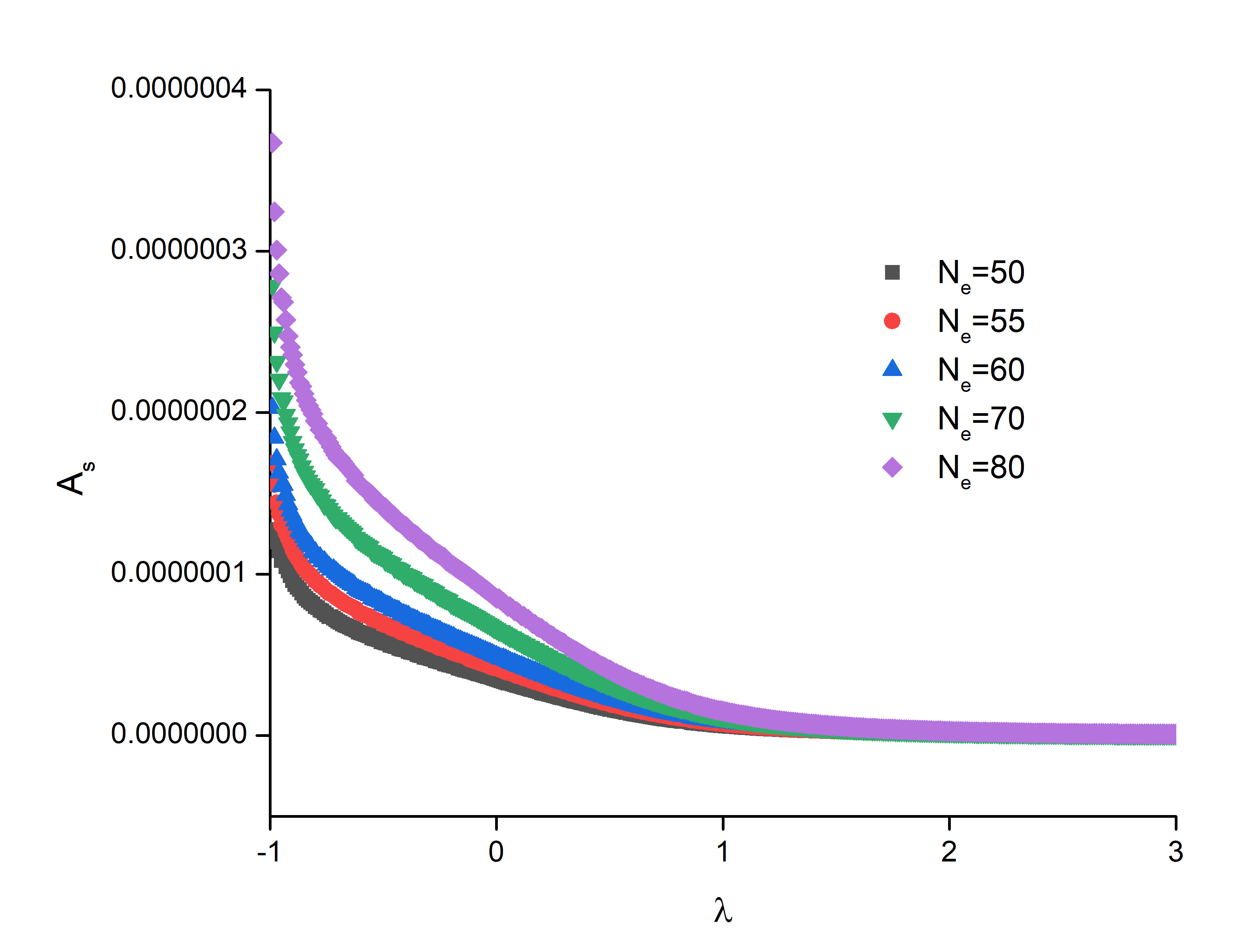}
    \caption{}
    \label{As}
\end{subfigure}
\begin{subfigure}{.47\textwidth}
    \centering
    \includegraphics[width=.8\linewidth]{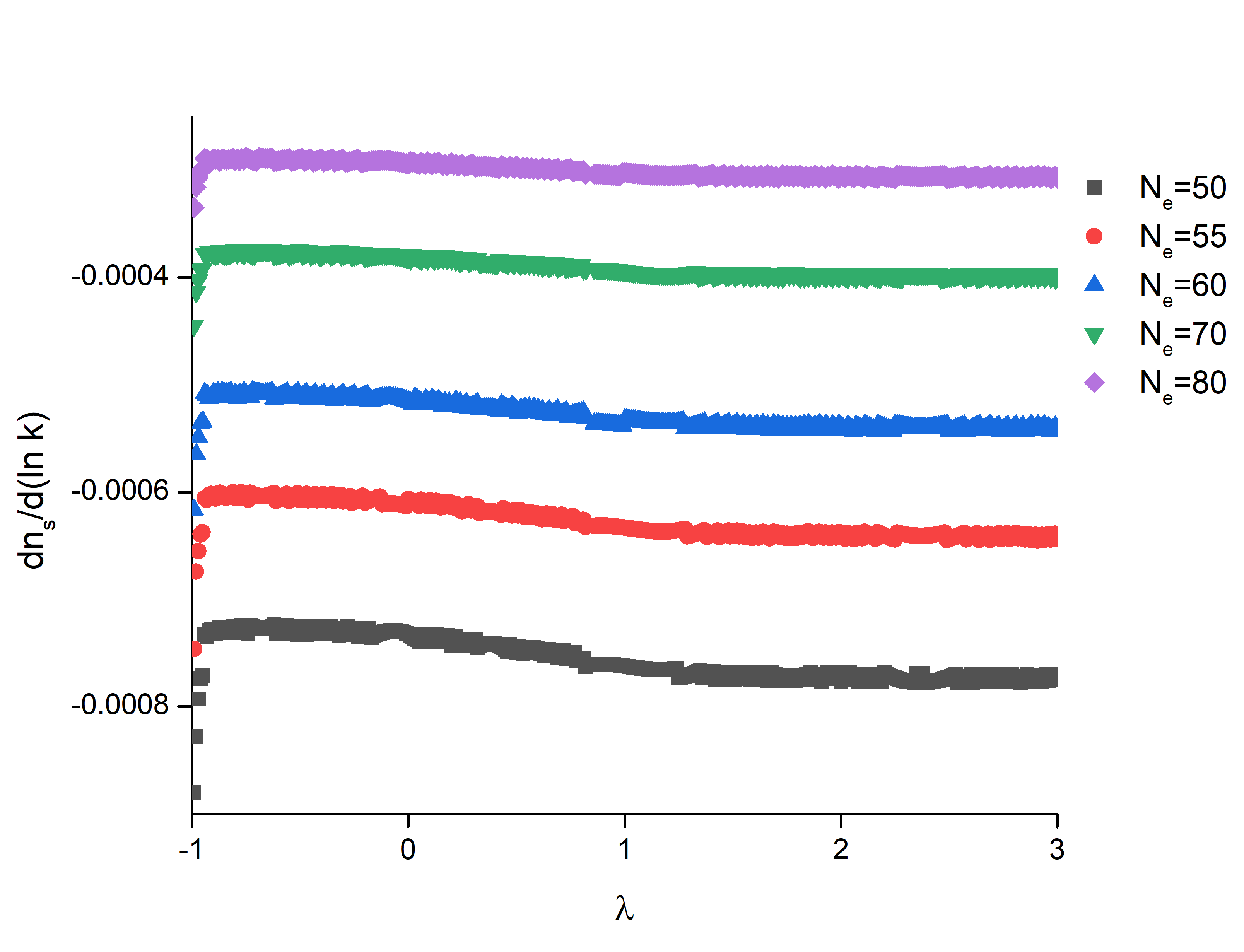}
    \caption{}
    \label{dns}
\end{subfigure}
    \caption{The dependence of a) $n_s$, b) r, c) $A_s$ and d) $\frac{dn_s}{d(lnk)}$ on $\lambda$}
    \label{fig:onefieldgraphs}
\end{figure}

As it is shown in figure \ref{fig:onefieldgraphs} , for $\lambda>1$, the quantities remain unchanged, after the dramatically change from $\lambda \approx -1$ to $\lambda =1$. Interestingly, for very small $\lambda$ near -1, the quantities vary rapidly.

 Compare to the observed values at 68\% confidence limit are $n_s = 0.9649 \pm 0.0042$, $dn_s/d \ln k = -0.0045 \pm 0.0064$, $A_s \approx 2.2 \times 10^{-9}$ and the upper bound for r with 95\% confidence limit is $r<0.063$. The table $\ref{tab:onefield}$ show that the prediction from the one-field potential fits the observation, except that the scalar perturbation amplitude is a little bit higher \cite{planck_collaboration_planck_2020}.

\subsection{Inflation with a two-fields potential}

 The general action for two fields inflation:
\begin{equation}
    S_4[\hat{g}_{AB}, \chi, f] =   \int d^4x\sqrt{-\hat{g}_4} \left( \frac{M^2_{Pl}}{2} \hat{R} - \frac{1}{2} \hat{g}^{\alpha\beta}\partial_\alpha \hat{\chi}\partial_\beta \hat{\chi} - \frac{1}{2} \hat{g}^{\alpha\beta} \partial_\alpha \hat{f} \partial_\beta \hat{f} - M^2_{Pl} V(\hat{\chi}, \hat{f}) \right).\\
\end{equation}

In this inflation model, the potential $V(\hat{\chi}, \hat{f})$ takes the form:
\begin{equation}
    M^{-4}_{Pl} V(\hat{\chi}, \hat{f}) = \left[ a^{-2}\left(1-e^{\frac{-6\hat{f}}{\sqrt{42}{M_{Pl}}}}\right)^\frac{4}{3} + 2\tilde{\Lambda}_8 e^{\frac{-8\hat{f}}{\sqrt{42}{M_{Pl}}}} \right] e^{\frac{-4\hat{\chi}}{2\sqrt{3}{M_{Pl}}}} - 2e^{\frac{-6\hat{\chi}}{2\sqrt{3}{M_{Pl}}}} + F^2 e^{\frac{-12\hat{\chi}}{2\sqrt{3}{M_{Pl}}}},
\end{equation}
with canonical fields $\hat{f}$ and $\hat{\chi}$ instead of f and $\chi$:
\begin{align}
        \hat{\chi} &= 2\sqrt{3}M_{Pl} \chi,\\
        \hat{f} &= \sqrt{42}M_{Pl}.
\end{align}

The calculation of inflationary quantities with two-fields potential is difficult due to the complexity of the potential itself. Then, it is reasonable to switch the attention to the Hubble parameter, and the dependence of both fields in inflationary time t. The Hubble parameter evolved in time by the Friedmann equations:
\begin{align}
    \label{H2} H^2 &= \frac{1}{2 M_{Pl}^2} \left( \frac{1}{2} \dot{f}^2 + \frac{1}{2}\dot{\chi}^2 + V(f, \chi) \right),\\
    \label{dotH} \dot{H} &= -\frac{1}{2M_{Pl}^2} \left( \dot{f}^2 + \dot{\chi}^2 \right).
\end{align}

The defined slow roll parameter $\epsilon=-\frac{\dot{H}}{H^2}$ and $\eta = -\frac{1}{H} \frac{\ddot{H}}{\dot{H}}$, they take the new forms in this two-fields formalism: \cite{asadi_reheating_2019, gashti_two-field_2021}
\begin{align}
        \epsilon &= \frac{3(\dot{f}^2 + \dot{\chi}^2)}{\dot{f}^2 + \dot{\chi}^2 + 2V(f,\chi)},\\
        \eta &= -\frac{2(\dot{f}\ddot{f} + \dot{\chi}\ddot{\chi})}{H(\dot{f}^2 + \dot{\chi}^2)}.
\end{align}

With the last parameter $\zeta^2$, we define it as:
\begin{equation}
    \zeta^2 = -4M_{Pl}^4 \frac{H^{(3)}\times H^{(2)}}{H^4}
\end{equation}
The potential in this model have a high complexity form, therefore calculating the Hubble parameter is impossible. Therefore, it is necessary to approximate the Hubble parameter into a much simpler form. In some previous works, the Hubble parameter is separable, so that the Hamilton-Jacobi formalism can be applied: \cite{asadi_reheating_2019, gashti_two-field_2021}
\begin{equation}
    H = H_0 + H_1 f + H_2 \chi.
\end{equation}

But this form of Hubble parameter does not reflex the nature of the potential of this work, therefore, we can consider some other forms of the Hubble parameter that can still apply the Hamilton-Jacobi formalism:
\begin{align*}
    H &= H_0 + H_1 f \chi + H_2 \chi,\\
    \textrm{or} \quad H &= H_0 + H_1 \frac{1}{\chi} f + H_2 \chi,\\
    \textrm{or} \quad H &= H_0 + H_1 \frac{1}{\chi} f + H_2 \frac{1}{\chi}, \textrm{etc.}
\end{align*}

In order to choose the most suitable form, we have used the 4th-order Runge-Kutta method to numerically calculate the $\epsilon$ and $\eta$  and compare them to the experimental data. Fortunately, the most suitable model is the simplest one, and this form also can be solved analytically:
\begin{equation}\label{HHamilJacob}
    H = H_0 + H_1 f \chi + H_2 \chi,
\end{equation}

There are three parameters in this approximation. The first one is $H_0$ which can be calculated from the Friedmann equation (\ref{H2}). $H_1$ and $H_2$ should be negative due to the nature of the potential and because the second term of the sum contains the combination of two field, $|H_1| < |H_2|$ and both $|H_1|$ and $ |H_2|$ must be extremely small.

Next we apply the Hamilton-Jacobi formalism, we know that the Hubble parameter depends on two variables f and $\chi$: $H(f, \chi)$, its first derivative:
\begin{equation}
    dH(f, \chi) = \frac{\partial H}{\partial f} df + \frac{\partial H}{\partial \chi} d\chi,
\end{equation}
or
\begin{equation}
    \dot{H} = \frac{\partial H}{\partial f} \dot{f} + \frac{\partial H}{\partial \chi} \dot{\chi}.
\end{equation}

Compares to the Friemann equation \ref{dotH}:
\begin{equation}
    \dot{H} = -\frac{1}{2M_{Pl}^2}\dot{f}\dot{f} -\frac{1}{2M_{Pl}^2} \dot{\chi}\dot{\chi},
\end{equation}
we have the system of PDEs:
\begin{equation}
    \begin{cases}
    \dot{f} = -2M_{Pl}^2 \frac{\partial H}{\partial f},\\
    \dot{\chi} = -2M_{Pl}^2 \frac{\partial H}{\partial \chi}.
    \end{cases}
\end{equation}
Using the form of Hubble parameter (\ref{HHamilJacob}), the system becomes:
\begin{equation}
    \begin{cases}
    \dot{f} = -2M_{Pl}^2 H_1 \chi,\\
     \dot{\chi} = -2M_{Pl}^2 (H_1 f + H_2 ).
    \end{cases}
\end{equation}

As we mentioned above, this system of PDEs were analytically solvable, the results are:
\begin{equation}
    \begin{cases}
    f = e^{-2M_{Pl}^2 H_1 t + f_0}-\frac{H_2}{H_1},\\
    \chi = e^{-2M_{Pl}^2 H_1 t + \chi_0},
    \end{cases}
\end{equation}
where $f_0$ and $\chi_0$ are the initial phrases of two fields, they were the initial values in Runge-Kutta methods we used above. These initial phrases also affected the parameter $H_0$. For convenience, we reassigned parameters $H_1$ and $H_2$:
\begin{align}
    \omega =& -2M_{Pl}^2 H_1,\\
    \delta =& \frac{H_2}{H_1}.
\end{align}

All initial parameters including $F^2$, a and $\lambda$ are fixed at $F^2 = 10^6$, $a=10$ and $\lambda = 0.01$. It is recognizable that the changing of a or $\lambda$ do not alter much the result. On the other hand, $F^2$ play an enormous role in results.

Of course both $\omega$ and $\delta$ must be positive. In this work, we assumed that $\chi_0 = 0$ and $f_0 = \ln{\delta}$ because it is important that the inflaton field f is not negative at the early inflation. 
After optimisation of the parameters $\omega$ and $\delta$, we find that at the values $\omega = -6.9$ and $\delta = 0.9$. The values of each quantities at $N_e=60$ are: r=0.03393, $n_s = 0.96902$, $dn_s/d \ln k = -0.40566$ and $A_s = 1.761\times 10^{-8}$. Compare to the observed values at 68\% confidence limit are $n_s = 0.9649 \pm 0.0042$, $dn_s/d \ln k = -0.0045 \pm 0.0064$, $A_s \approx 2.2 \times 10^{-9}$ and the upper bound for r with 95\% confidence limit is $r<0.063$ \cite{planck_collaboration_planck_2020}. It can be verified that both scalar-to-tensor ratio r and spectral index fit the observation, while scalar perturbation amplitude was higher and running index is much higher. It can be concluded that the two-fields potential model can only partial predict the observation and the stabilization of the moduli $\chi$ is essential.

The dependence of each observable quantities on two parameters $\omega$ and $\delta$ at $N_e = 60$:
\begin{figure}[h] 
    \begin{subfigure}{.47\textwidth}
    \centering
    \includegraphics[width=.8\linewidth]{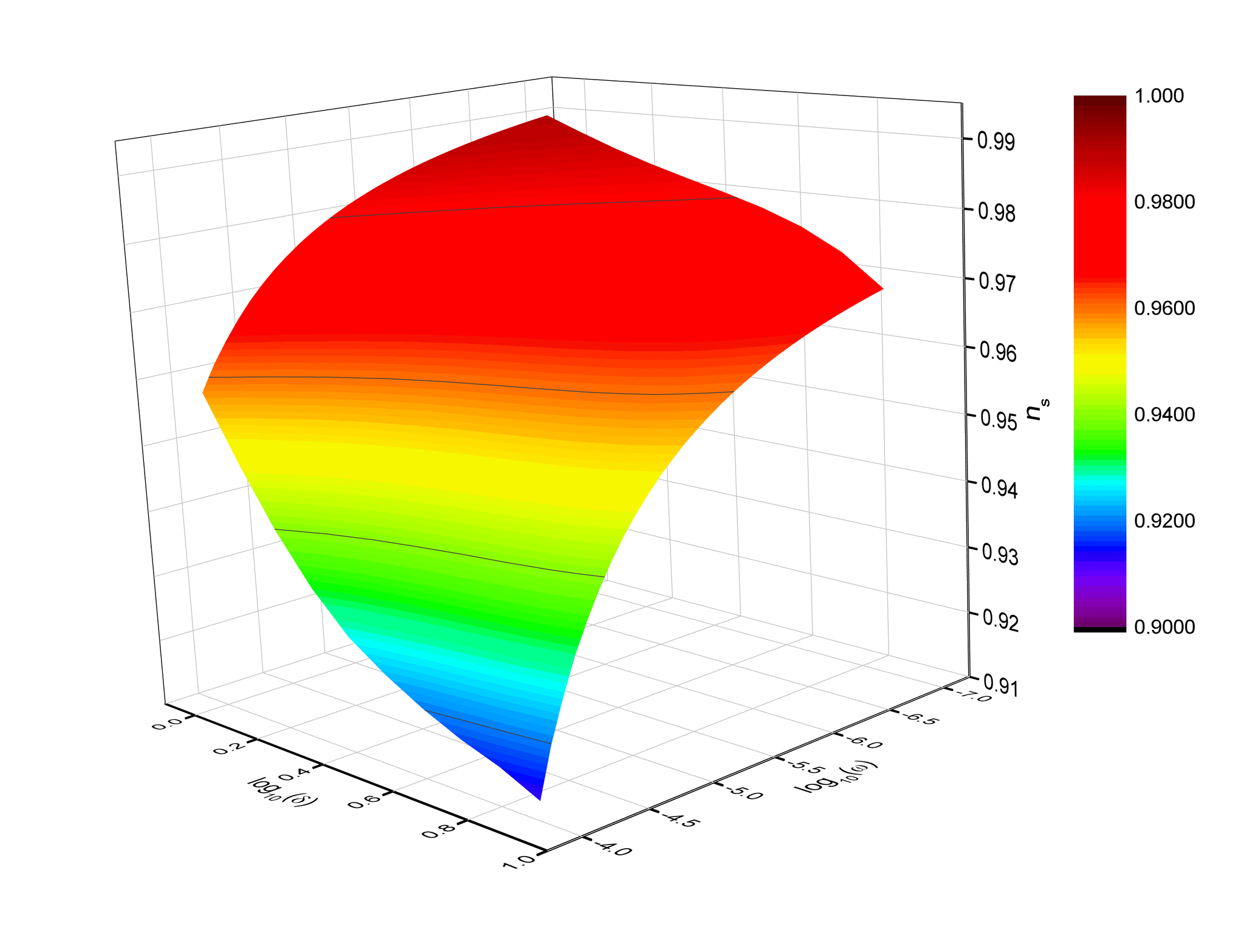}
    \caption{}
    \label{ns}
\end{subfigure}
\begin{subfigure}{.47\textwidth}
    \centering
    \includegraphics[width=.8\linewidth]{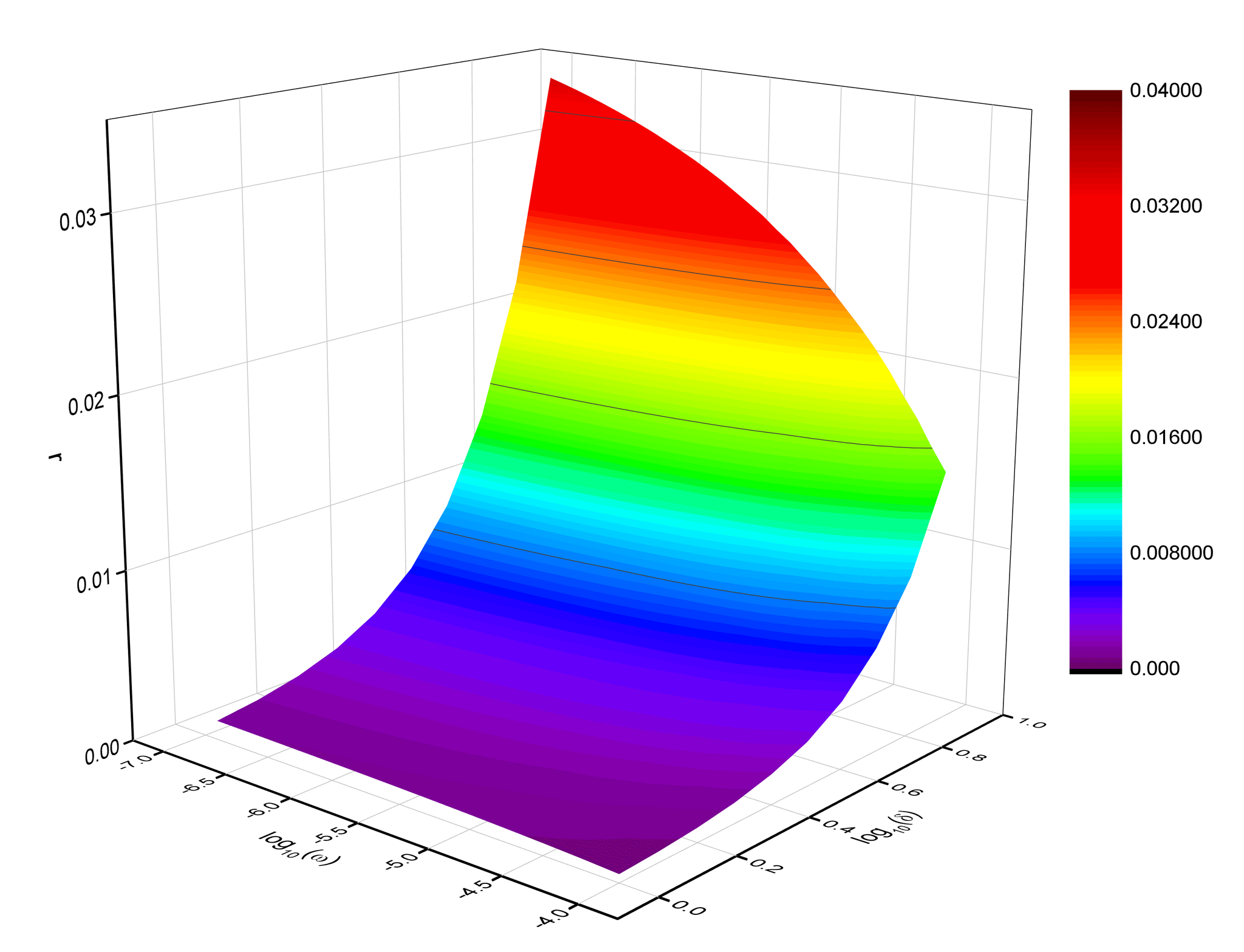}
    \caption{}
    \label{r}
\end{subfigure}
\begin{subfigure}{.47\textwidth}
    \centering
    \includegraphics[width=.8\linewidth]{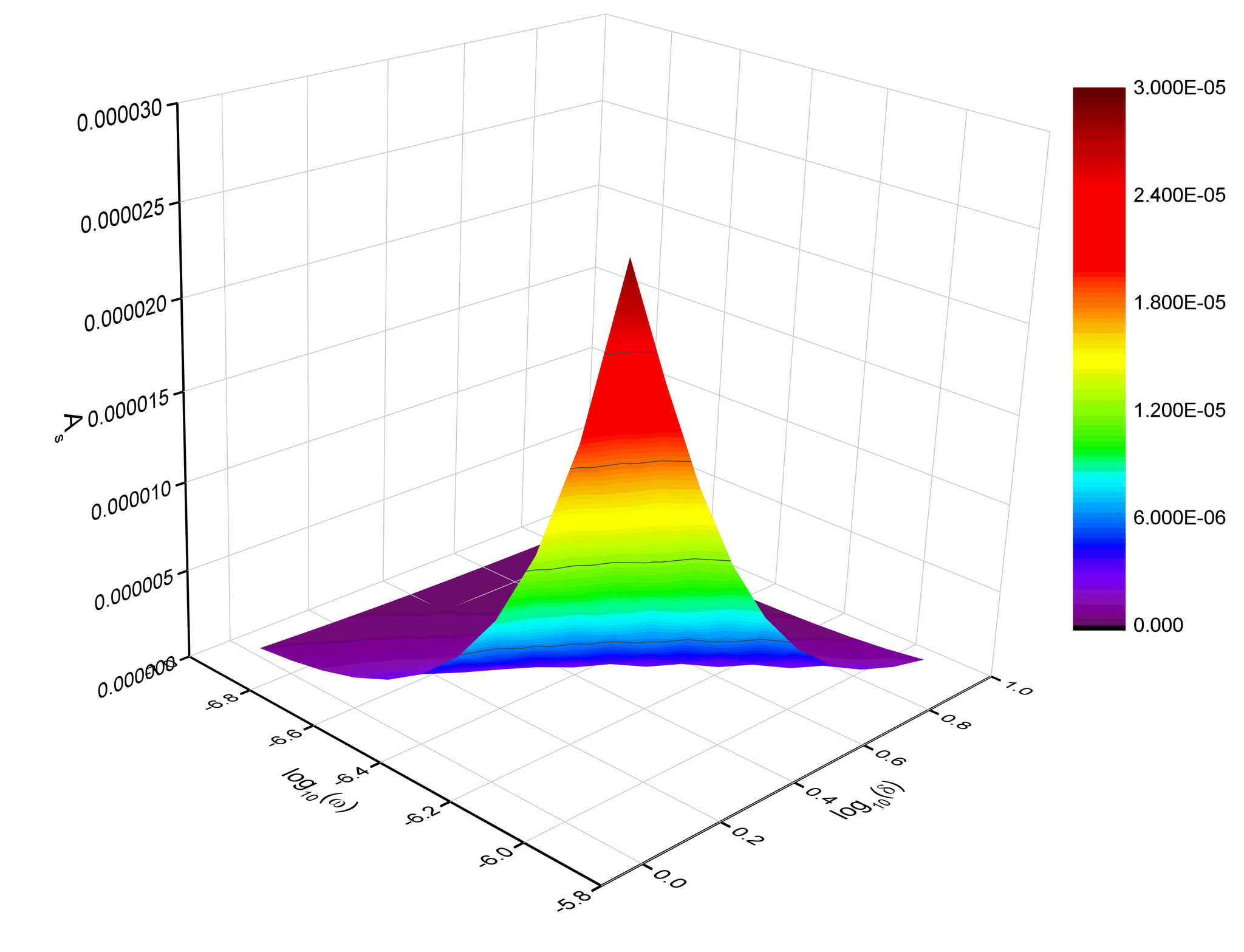}
    \caption{}
    \label{As}
\end{subfigure}
\begin{subfigure}{.47\textwidth}
    \centering
    \includegraphics[width=.8\linewidth]{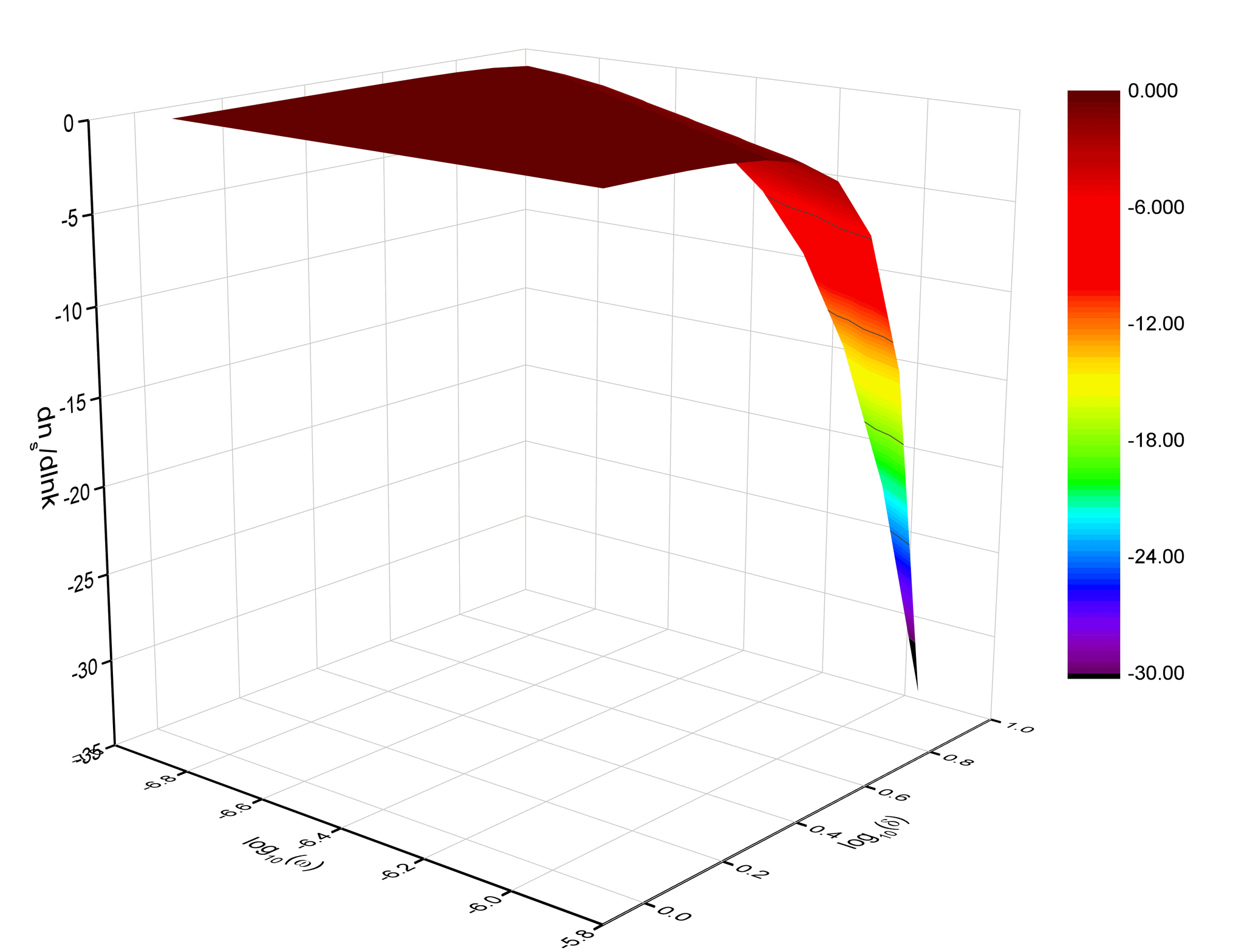}
    \caption{}
    \label{dns}
\end{subfigure}
    \caption{The dependence of a) $n_s$, b) r, c) $A_s$ and d) $\frac{dn_s}{d(lnk)}$ on $\omega$ and $\delta$}
    \label{fig:twofieldgraphs}
\end{figure}
At the the optimisation of parameters, the optimal of parameters are $\omega = -6.9$ and $\delta=0.9$. If changing the e-folding numbers, the quantities change through very simple patterns. The changes are obvious but not dramatic:
\begin{figure}[h] 
    \begin{subfigure}{.47\textwidth}
    \centering
    \includegraphics[width=.8\linewidth]{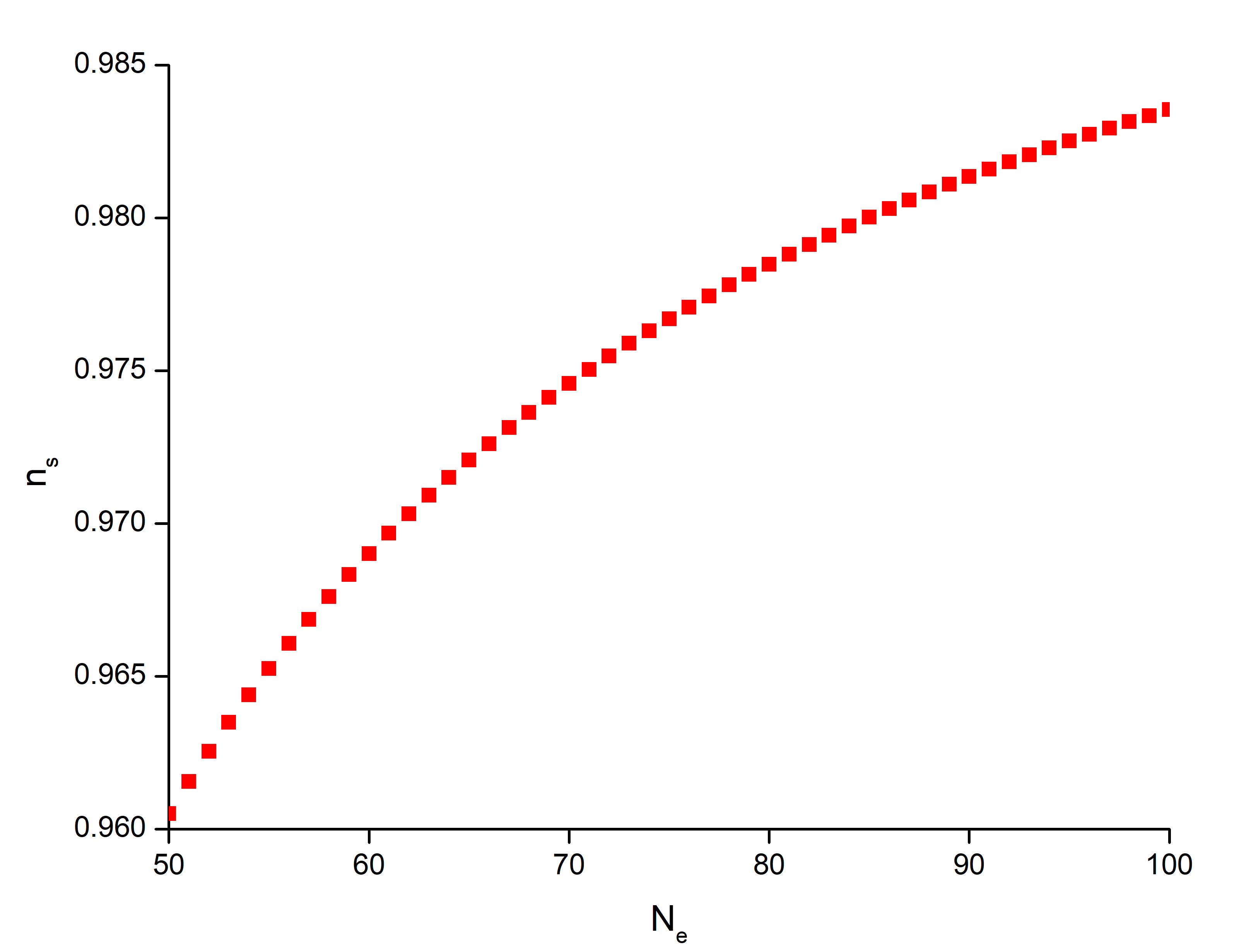}
    \caption{}
    \label{ns}
\end{subfigure}
\begin{subfigure}{.47\textwidth}
    \centering
    \includegraphics[width=.8\linewidth]{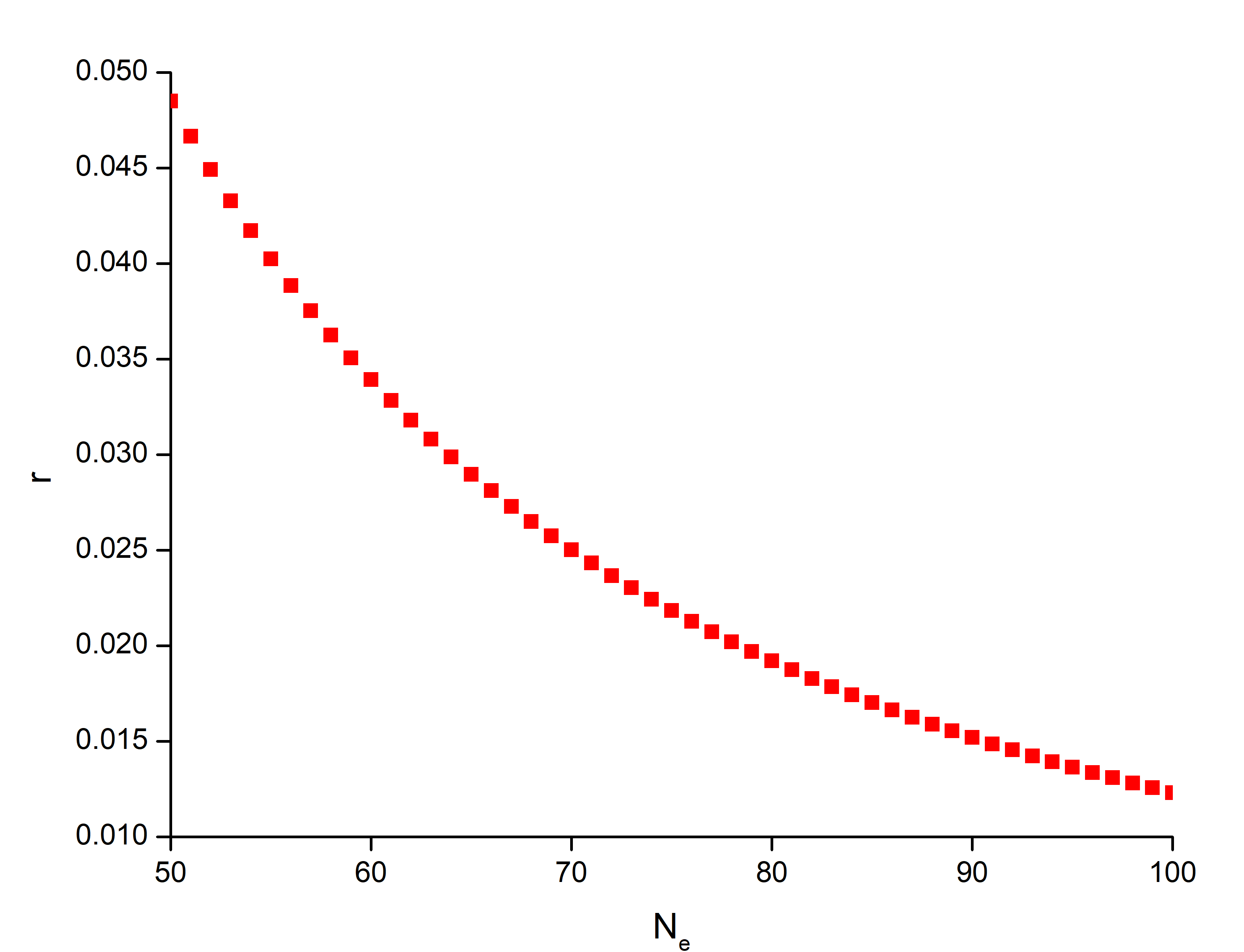}
    \caption{}
    \label{r}
\end{subfigure}
\begin{subfigure}{.47\textwidth}
    \centering
    \includegraphics[width=.8\linewidth]{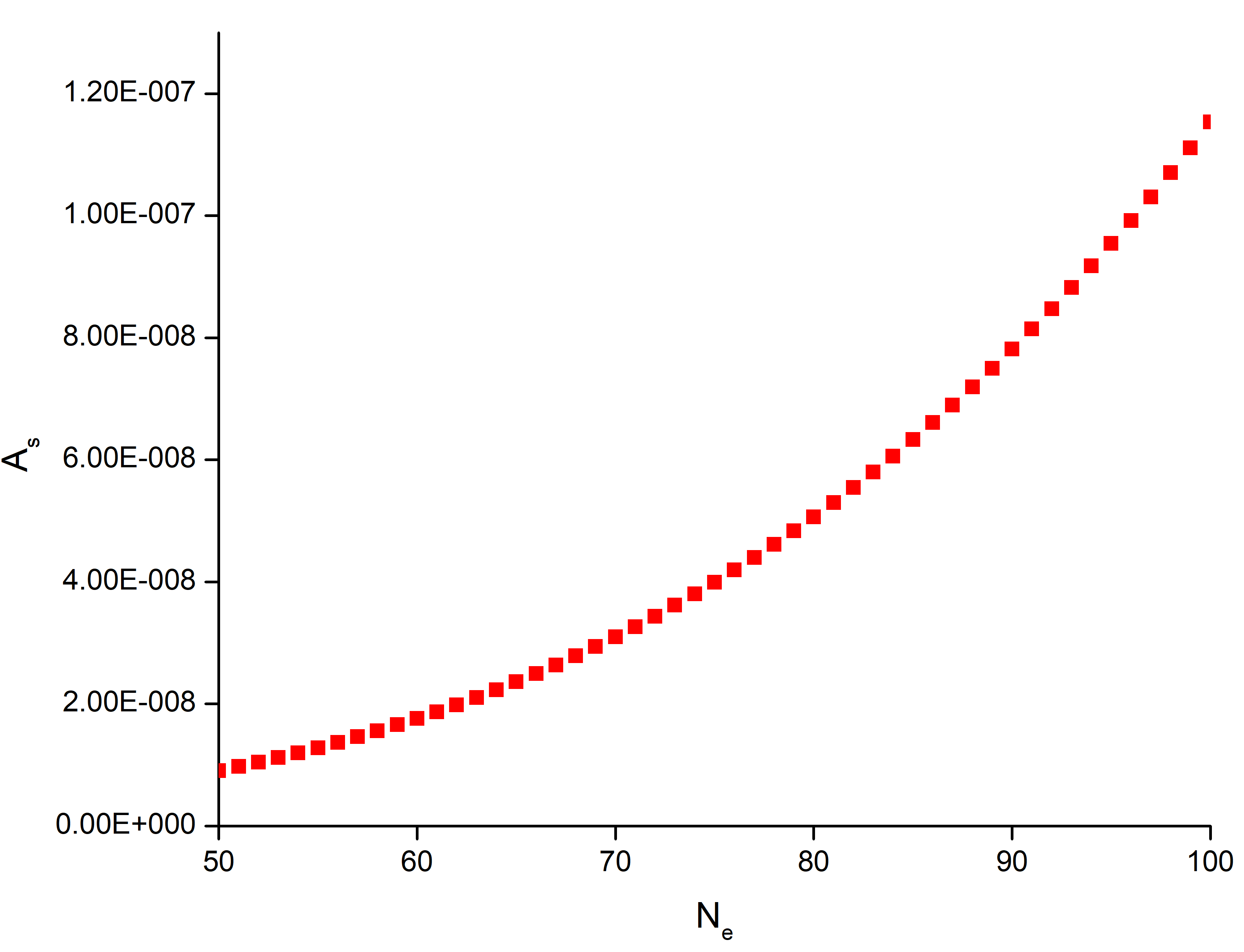}
    \caption{}
    \label{As}
\end{subfigure}
\begin{subfigure}{.47\textwidth}
    \centering
    \includegraphics[width=.8\linewidth]{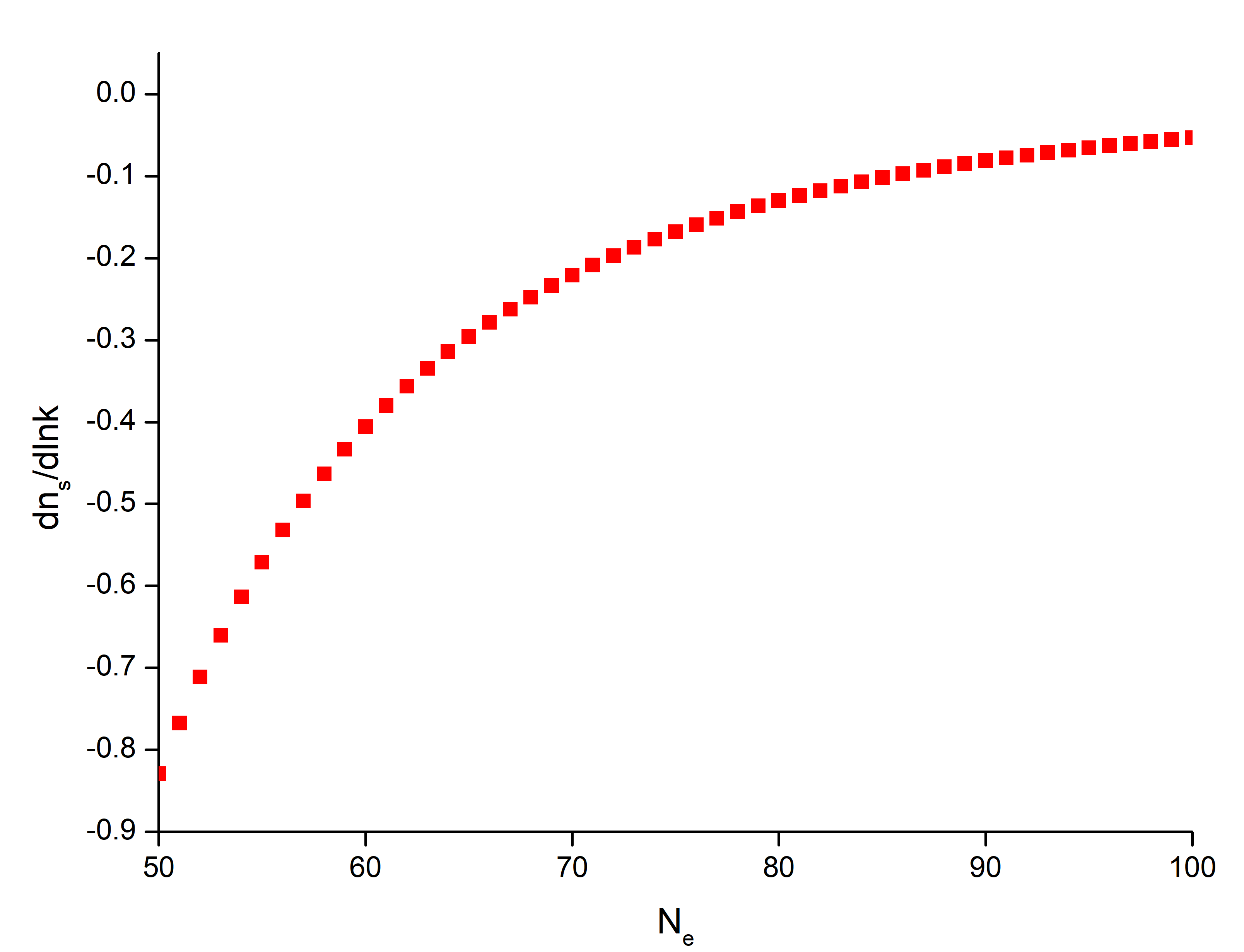}
    \caption{}
    \label{dns}
\end{subfigure}
    \caption{The dependence of a) $n_s$, b) r, c) $A_s$ and d) $\frac{dn_s}{d(lnk)}$ at optimal values of parameters $\omega =-6.9$ and $\delta = 0.9$ on e-folding numbers}
    \label{fig:twofieldNegraphs}
\end{figure}

\section{Swampland dS conjectures}
\subsection{de Sitter conjectures with single-field potential}

For N = 60 and $\lambda$ = 0.01, the field is at: $\hat{f} = 0.7200252x\sqrt{42}\times M_{Pl}$. Applying the original dS conjectures, if V is a single field inflation potential, the first conjecture is just a first derivative and the second one is a second derivative.

The defined slow-roll parameters:
\begin{align}
    \epsilon_V = \frac{M_p^2}{2} \left( \frac{V'}{V} \right)^2\\
    \eta_V = M^2_{Pl} \frac{V''}{V},
\end{align}
substitute into refined de Sitter conjectures, the Swampland conjectures written in slow-roll parameters:
\begin{equation}
    \sqrt{2\epsilon_V} \geq c_1 \quad \textrm{or} \quad \eta_V. \leq -c_2
\end{equation}

For $\hat{f} = 0.723437\sqrt{42}M_{Pl}$, the slow-roll parameters:
\begin{equation}
    \epsilon_V = 0.000155, \quad \eta_V = -0.0162,
\end{equation}
and the bounds for swampland parameters:
\begin{align}
    c_1 \leq 0.018,\\
    c_2 \leq 0.016.
\end{align}

Both are not of order 1, thus, de Sitter conjectures are not satisfied with this potential. Therefore, we keep studying the further refined de Sitter conjecture.

For different values of cosmological constant - $\lambda$, the boundary on $c_1$ and $c_2$ vary as:
\begin{figure}[h]
\begin{subfigure}{.47\textwidth}
    \centering
    \includegraphics[width=.8\linewidth]{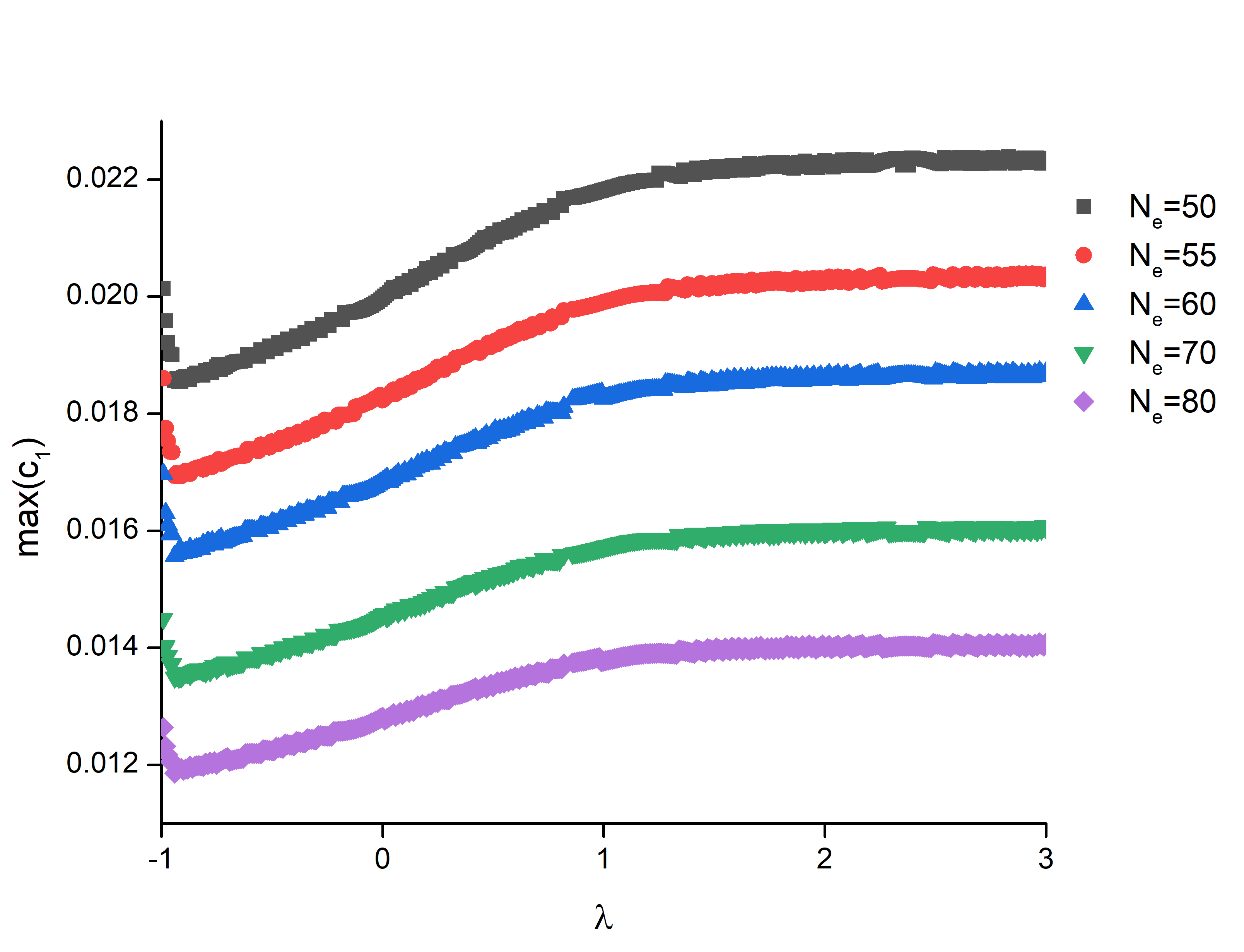}
    \caption{$c_1$}
    \label{c1}
\end{subfigure}
\begin{subfigure}{.47\textwidth}
    \centering
    \includegraphics[width=.8\linewidth]{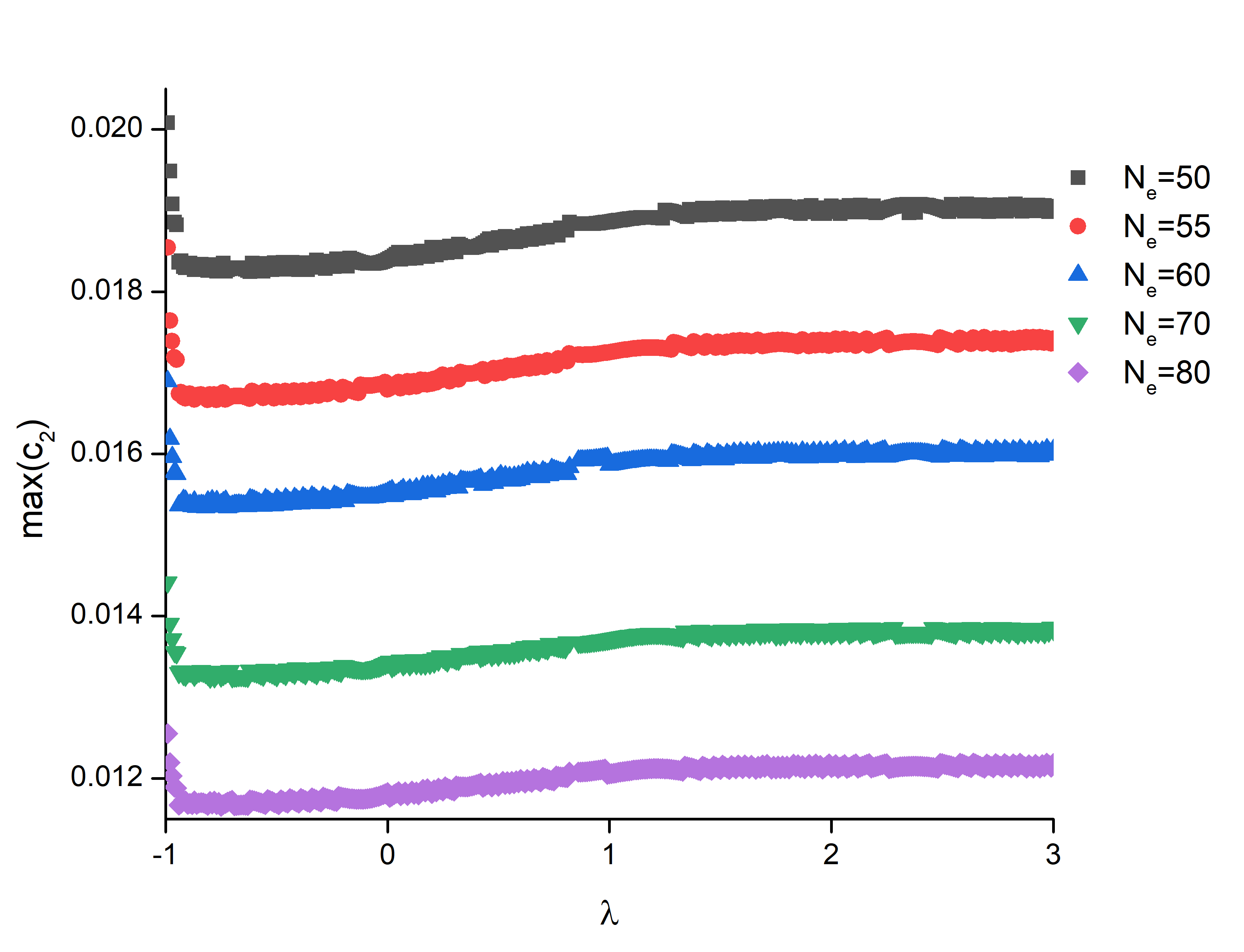}
    \caption{$c_2$}
    \label{c2}
\end{subfigure}
    \caption{The variation of the boundary of a) $c_1$ and b) $c_2$ depends on $\lambda$}
    \label{fig:my_label}
\end{figure}
We can rewrite the further refined de Sitter conjecture in slow-roll parameters as:
\begin{equation}
    (2\epsilon_V)^\frac{q}{2} - a\eta_V \geq b.
\end{equation}
Substitute the values of calculated slow-roll parameters, the result:
\begin{equation}
    0.017^q + 0.016a \geq 1-a,
\end{equation}
or
\begin{align}
\label{aqswlp} \frac{1}{0.017}(1-0.016^q) -a < 0.
\end{align}

It is easy to find a $q > 2$ and an "a" that satisfy the above relation. Indeed, for any $a < 0.984$, there is always a "q" that satisfy. Therefore, it is possible that the theory relates to this potential is an effective quantum gravity theory.

\begin{figure}[h]
    \centering
    \includegraphics[width = 8cm]{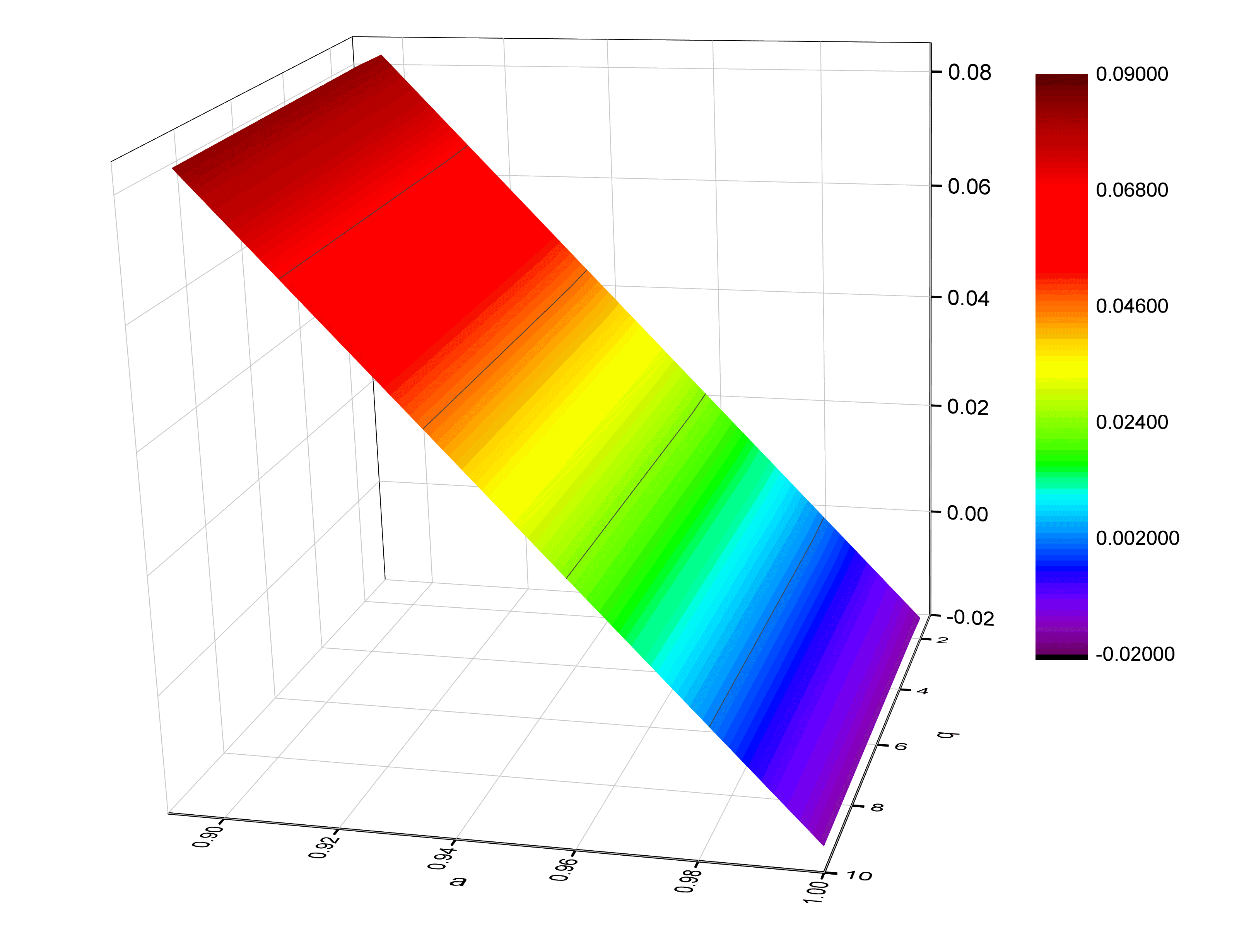}
    \caption{For $a > 0.984$, the relation \ref{aqswlp} is satisfied.}
    \label{fig:my_label}
\end{figure}

\subsection{de Sitter conjectures with Two-fields potential}

Applying the gradients of the potentials to de Sitter conjecture, the boundary on $c_1$:
\begin{align*}
    c_1 \leq \frac{|\nabla V(f_0, \chi_c)|\times M_P}{V(f_0, \chi_c)} = \frac{\sqrt{\partial_f V(f_0,\chi_c)^2 + \partial_x V(f_0,\chi_c)^2 }\times M_P}{V(f_0,\chi_c)}.
\end{align*}
Or $c_1 \leq 1.1464$. The result showed that $c_1$ can satisfy the condition of order one. This implies that the two-fields potential possibly is in the landscape.
Next, we study the refined de Sitter conjecture, put on the values ($f_0,\chi_c$) and receive the minimum of the eigenvalues matrix is $-8.7208 \times 10^{-22} $, the boundary on $c_2$ is:
\begin{equation}
    c_2 < -(-8.7208 \times 10^{-22}) \times M_P^2/V(f_0,\chi_c) = 0.0110.
\end{equation}
Therefore, the potential does not satisfy the refined de Sitter conjecture.

Studying the further refined de Sitter conjecture:
\begin{align}
    1.1464^q + 0.0110a \geq 1-a,\\
    \textrm{or} \quad \frac{1}{1.0110}(1-1.1464^q) \leq a < 1,\\
    \textrm{or} \quad \label{aqswampland} \frac{1}{1.0110}(1-1.1464^q) - a < 0.
\end{align}

Since $\frac{1}{1.0110}(1-1.1464^q)$ it can be concluded that for any a and q, the relation is satisfied.

\begin{figure}[h]
    \centering
    \includegraphics[width = 10cm]{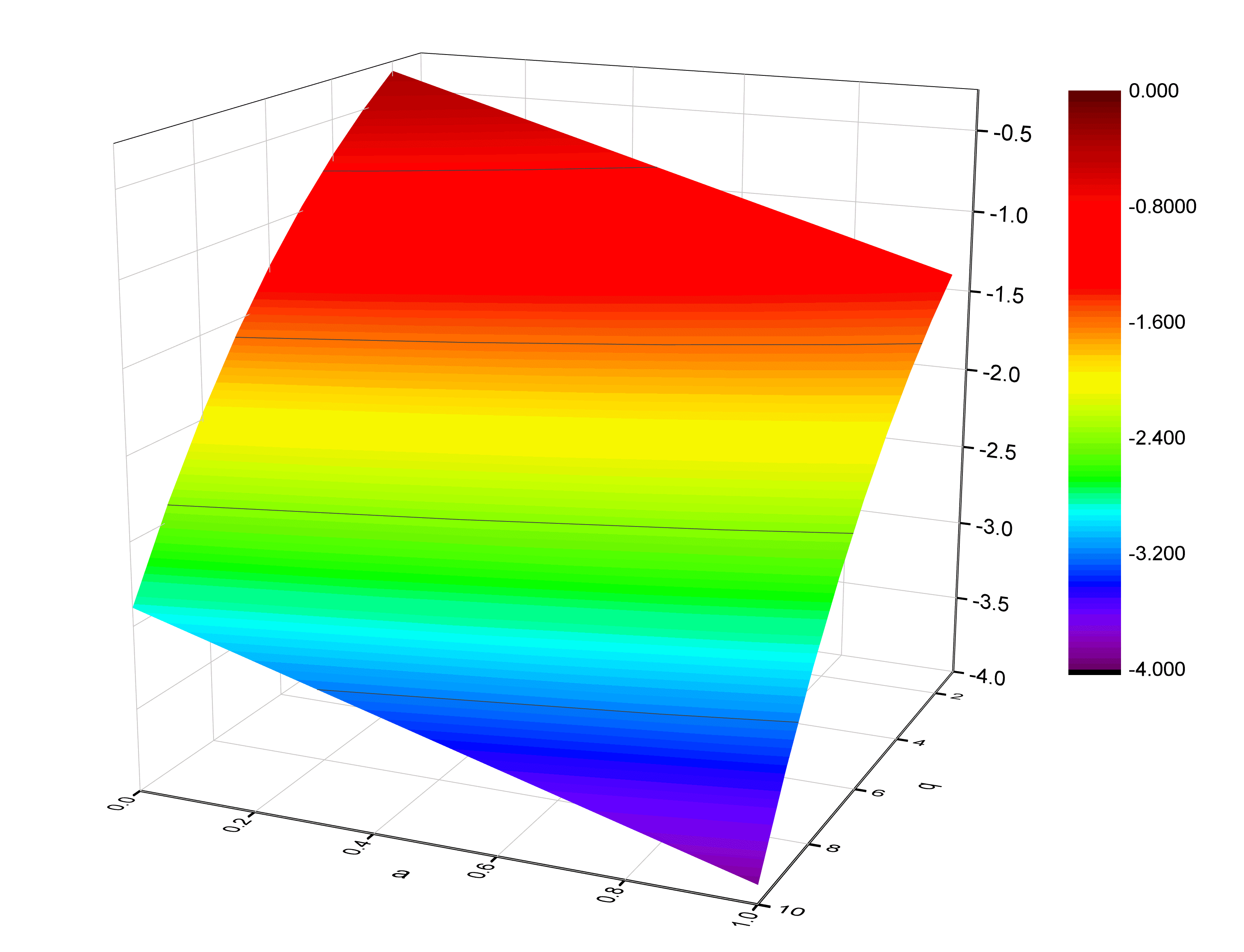}
    \caption{For any $0<a<1$ and $q>2$, the relation \ref{aqswampland} satisfy}
    \label{fig:my_label}
\end{figure}

\section{Conclusion}

In this work, we had studied the single-field potential and two-fields potential obtained from the modified gravity in 8-dimensional spacetime. The two-fields potential, included an inflaton field f and a radion field $\chi$, arised from the compactification from 8-dimensional spacetime on a 4-sphere with a warp factor and a flux. The single-field potential was obtained by the stabilization of moduli $\chi$. Four observable quantities included the scalar-to-tensor ratio r, spectral index $n_s$, running index $dn_s/d\ln k$ and the scalar amplitude $A_s$. The prediction from the single-field potential fits the experimental data better. The two-fields potential gave a prediction that also fits the measurement but with strict constraint on its parameters and e-folding numbers.

On the other hand, we also studied whether the potentials are in the Landscape of quantum gravity by de Sitter conjectures. Unfortunately, the single-field potential does not satisfied he original de Sitter and the refined de Sitter conjecture as well. But if we put consideration into the new further refined de Sitter conjecture, this potential could satisfy the conjecture within a range of parameters a and q. In contrast, the two-fields potential satisfy the original de Sitter conjecture. Moreover, it satisfied the further refined conjecture with all values of q and a. 

\iffalse

\fi 

\bibliography{Thesis}

@ARTICLE{randall_large_1999,
	title = "{Large Mass Hierarchy from a Small Extra Dimension}",
	volume = {83},
	issn = {0031-9007, 1079-7114},
	url = {https://link.aps.org/doi/10.1103/PhysRevLett.83.3370}, year = {1999},
	doi = {10.1103/PhysRevLett.83.3370},
	pages = {3370--3373},
	number = {17},
	journal = {Physical Review Letters},
	shortjournal = {Phys. Rev. Lett.},
	author = {Randall, Lisa and Sundrum, Raman},
	urldate = {2022-02-21},
	date = {1999-10-25},
	langid = {english},
}

@ARTICLE{douglas_flux_2007,
	title = {Flux compactification},
	volume = {79}, year = {2017},
	url = {https://link.aps.org/doi/10.1103/RevModPhys.79.733},
	doi = {10.1103/RevModPhys.79.733},
	pages = {733--796},
	number = {2},
	journal = {Rev. Mod. Phys.},
	author = {Douglas, Michael R. and Kachru, Shamit},
	date = {2007-05},
	note = {Publisher: American Physical Society},
}

@ARTICLE{nakada_inflation_2017,
	title = {Inflation from higher dimensions},
	volume = {96}, year = {2017},
	issn = {2470-0010, 2470-0029},
	doi = {10.1103/PhysRevD.96.123530},
	pages = {123530},
	number = {12},
	journal = {Physical Review D},
	shortjournal = {Phys. Rev. D},
	author = {Nakada, Hiroshi and Ketov, Sergei V.},
	urldate = {2022-07-07},
	date = {2017-12-26},
	langid = {english},
	
}

@ARTICLE{ooguri_distance_2019,
	title ="{Distance and de Sitter Conjectures on the Swampland}",
	volume = {788}, year = {2019},
	issn = {03702693},
	doi = {10.1016/j.physletb.2018.11.018},
	pages = {180--184},
	journal = {Physics Letters B},
	shortjournal = {Physics Letters B},
	author = {Ooguri, Hirosi and Palti, Eran and Shiu, Gary and Vafa, Cumrun},
	urldate = {2022-07-07},
	date = {2019-01},
	langid = {english},
}

@misc{vafa_string_2005,
	title = {The String Landscape and the Swampland},
	number = {{arXiv}:hep-th/0509212},
	publisher = {{arXiv}},
	author = {Vafa, Cumrun},
	urldate = {2022-07-07},
	date = {2005-10-06},
	langid = {english},
}

@ARTICLE{linde_axions_1991,
	title = {Axions in inflationary cosmology},
	volume = {259},
	issn = {0370-2693}, year = {1991},
	url = {https://www.sciencedirect.com/science/article/pii/037026939190130I},
	doi = {https://doi.org/10.1016/0370-2693(91)90130-I},
	pages = {38--47},
	number = {1},
	journal = {Physics Letters B},
	author = {Linde, Andrei},
	date = {1991},
}

@ARTICLE{linde_chaotic_1983,
	title = {Chaotic inflation},
	volume = {129},
	issn = {0370-2693}, year = {1983},
	url = {https://www.sciencedirect.com/science/article/pii/0370269383908377},
	doi = {https://doi.org/10.1016/0370-2693(83)90837-7},
	pages = {177--181},
	number = {3},
	journal = {Physics Letters B},
	author = {Linde, A. D.},
	date = {1983},
}

@ARTICLE{linde_new_1982,
	title = {A {NEW} {INFLATIONARY} {UNIVERSE} {SCENARIO}: A {POSSIBLE} {SOLUTION} {OF} {THE} {HORIZON}, {FLATNESS}, {HOMOGENEITY}, {ISOTROPY} {AND} {PRIMORDIAL} {MONOPOLE} {PROBLEMS}},
	volume = {108},
	pages = {5}, year = {2012},
	number = {6},
	journal = {{PHYSICS} {LETTERS}},
	author = {Linde, A D},
	date = {1982},
	langid = {english},
}

@ARTICLE{linde_current_2005,
	title = {Current understanding of inflation},
	volume = {49}, year = {2005},
	issn = {13876473},
	url = {https://linkinghub.elsevier.com/retrieve/pii/S1387647305000047},
	doi = {10.1016/j.newar.2005.01.002},
	pages = {35--41},
	number = {2},
	journal = {New Astronomy Reviews},
	shortjournal = {New Astronomy Reviews},
	author = {Linde, Andrei},
	urldate = {2022-07-07},
	date = {2005-05},
	langid = {english},
}

@ARTICLE{barrow_premature_1988,
	title = "{The premature recollapse problem in closed inflationary universes}",
    year={1988},
	volume = {296},
	issn = {0550-3213},
	doi = {https://doi.org/10.1016/0550-3213(88)90040-5},
	pages = {697--709},
	number = {3},
	journal = {Nuclear Physics B},
	author = {Barrow, John D.},
	date = {1988},
}

@ARTICLE{planck_collaboration_planck_2020,
	title = {Planck 2018 results. X. Constraints on inflation},
	volume = {641}, year = {2020},
	issn = {0004-6361, 1432-0746},
	doi = {10.1051/0004-6361/201833887},
	pages = {A10},
	journal = {Astronomy \& Astrophysics},
	shortjournal = {A\&A},
	author = {Planck Collaboration and Akrami, et. al},
	urldate = {2022-07-07},
	date = {2020-09},
	langid = {english},
}

@ARTICLE{sotiriou_fr_2010,
	title = "{f(R) Theories Of Gravity}",
    year = "2010",
	volume = {82},
	issn = {0034-6861, 1539-0756},
	doi = {10.1103/RevModPhys.82.451},
        number = {1},
	pages = {451--497},
	journal = {Reviews of Modern Physics},
	shortjournal = {Rev. Mod. Phys.},
	author = {Sotiriou, Thomas P. and Faraoni, Valerio},
	urldate = {2022},
	date = {2010-03-01},
	langid = {english},

}

@ARTICLE{starobinsky_new_1980,
	title = {A new type of isotropic cosmological models without singularity},
	volume = {91},
    year = {1980},
	issn = {0370-2693},
	doi = {https://doi.org/10.1016/0370-2693(80)90670-X},
	pages = {99--102},
	number = {1},
	journal = {Physics Letters B},
	author = {Starobinsky, A. A.},
	date = {1980},
}

@ARTICLE{clifton_modified_2012,
	title = "{Modified Gravity and Cosmology}",
	volume = {513},
	issn = {03701573},
	doi = {10.1016/j.physrep.2012.01.001},
	pages = {1--189},
	number = {1},
	journal = {Physics Reports},
	shortjournal = {Physics Reports},
	author = {Clifton, Timothy and Ferreira, Pedro G. and Padilla, Antonio and Skordis, Constantinos},
	urldate = {2022-07-07},
	date = {2012-03},
    year = {2012},
	langid = {english},
}

@ARTICLE{danielsson_what_2018,
	title = {What if string theory has no de Sitter vacua?},
	volume = {27}, year = {2018},
	issn = {0218-2718, 1793-6594},
	doi = {10.1142/S0218271818300070},
	pages = {1830007},
	number = {12},
	journal = {International Journal of Modern Physics D},
	shortjournal = {Int. J. Mod. Phys. D},
	author = {Danielsson, Ulf H. and Van Riet, Thomas},
	urldate = {2022-06-18},
	date = {2018-09},
	langid = {english},
}

@ARTICLE{asadi_reheating_2019,
	title = {Reheating constraints on a two-field inflationary model},
	volume = {949},
	issn = {05503213},
    year = {2019},
	doi = {10.1016/j.nuclphysb.2019.114827},
	pages = {114827},
	journal = {Nuclear Physics B},
	shortjournal = {Nuclear Physics B},
	author = {Asadi, Kosar and Nozari, Kourosh},
	urldate = {2022-05-09},
	date = {2019-12},
	langid = {english},
}

@ARTICLE{gashti_two-field_2021,
	title = "{Two-Field Inflationary Model and Swampland de Sitter Conjecture}",
	journal = {{arXiv}:2111.06421 [gr-qc]},
	author = {Gashti, S. Noori},
	urldate = {2022-05-07},
	date = {2021-11-11},
    year = {2021},
	langid = {english},
}

@ARTICLE{andriot_further_2019,
	title = "{Further Refining the de Sitter Swampland Conjecture}", year = {2019},
	volume = {67},
	issn = {00158208},
	doi = {10.1002/prop.201800105},
	pages = {1800105},
	number = {1},
	journal = {Fortschritte der Physik},
	shortjournal = {Fortschr. Phys.},
	author = {Andriot, David and Roupec, Christoph},
	urldate = {2022-04-18},
	date = {2019-01},
	langid = {english},
}

@ARTICLE{palti_swampland_2019,
	title = "{The Swampland: Introduction and Review}", year = {2019},
	volume = {67},
	issn = {0015-8208, 1521-3978},
	doi = {10.1002/prop.201900037},
	shorttitle = {The Swampland},
	pages = {1900037},
	number = {6},
	journal = {Fortschritte der Physik},
	shortjournal = {Fortschr. Phys.},
	author = {Palti, Eran},
	urldate = {2022-03-28},
	date = {2019-06},
	langid = {english},
}

@ARTICLE{liu_higgs_2021,
	title = "{Higgs inflation and its extensions and the further refining {dS} swampland conjecture}",
	volume = {81}, year = {2021},
	issn = {1434-6044, 1434-6052},
	doi = {10.1140/epjc/s10052-021-09940-w},
	pages = {1122},
	number = {12},
	journal = {The European Physical Journal C},
	shortjournal = {Eur. Phys. J. C},
	author = {Liu, Yang},
	urldate = {2022-03-21},
	date = {2021-12},
	langid = {english},
}

@ARTICLE{otero_r_2017,
	title = "{$R + \alpha R^n$ inflation in higher-dimensional space-times}",
	volume = {2017},
	issn = {1029-8479}, year = {2019},
	url = {http://link.springer.com/10.1007/JHEP05(2017)058},
	doi = {10.1007/JHEP05(2017)058},
	pages = {58},
	number = {5},
	journal = {Journal of High Energy Physics},
	shortjournal = {J. High Energ. Phys.},
	author = {Otero, Santiago Pajón and Pedro, Francisco G. and Wieck, Clemens},
	urldate = {2022-03-17},
	date = {2017-05},
	langid = {english},
}

@ARTICLE{nam_implications_2021,
	title = {Implications for the hierarchy problem, inflation and geodesic motion from fiber fabric of spacetime},
	volume = {81},
	issn = {1434-6044, 1434-6052}, year = {2021},
	doi = {10.1140/epjc/s10052-021-09881-4},
	pages = {1102},
	number = {12},
	journal = {The European Physical Journal C},
	shortjournal = {Eur. Phys. J. C},
	author = {Nam, Cao H.},
	urldate = {2022-03-13},
	date = {2021-12},
	langid = {english},
}

@ARTICLE{ketov_modified_2019,
	title = "{Modified Gravity in Higher Dimensions, Flux Compactification, and Cosmological Inflation}",
	volume = {11},
    year = {2019},
	issn = {2073-8994},
	url = {https://www.mdpi.com/2073-8994/11/12/1528},
	doi = {10.3390/sym11121528},
	pages = {1528},
	number = {12},
	journal = {Symmetry},
	shortjournal = {Symmetry},
	author = {Ketov, Sergei V.},
	urldate = {2022-02-25},
	date = {2019-12-17},
	langid = {english},
}

@ARTICLE{ketov_inflation_2017,
	title = "{Inflation from $(R + \gamma R^n − 2 \Lambda)$ gravity in higher dimensions}",
	volume = {95},
	issn = {2470-0010, 2470-0029},
	url = {http://link.aps.org/doi/10.1103/PhysRevD.95.103507}, year = {2017},
	doi = {10.1103/PhysRevD.95.103507},
	pages = {103507},
	number = {10},
	journal = {Physical Review D},
	shortjournal = {Phys. Rev. D},
	author = {Ketov, Sergei V. and Nakada, Hiroshi},
	urldate = {2022-02-21},
	date = {2017-05-15},
	langid = {english},
}

\end{document}